\documentclass{article}
\usepackage[numbers,sort&compress]{natbib}
\usepackage{amsmath,graphicx,mlspconf}

\title{Learning Signal Representations for EEG Cross-Subject Channel Selection and Trial Classification}
\name{Michela C. Massi $^{a,b}$ and Francesca Ieva $^{a,b,c}$}
\address{$^{a}$ MOX - Dept. of Mathematics, Politecnico di Milano\\
$^{b}$ CADS - Human Technopole\\
$^{c}$ CHRP - Bicocca University\\
\thanks{\noindent \texttt{$\{$michelacarlotta.massi,francesca.ieva$\}$@polimi.it}
}
}

\begin{document}

\maketitle

\begin{abstract}
EEG is a non-invasive powerful system that finds applications in several domains and research areas. Most EEG systems are multi-channel in nature, but multiple channels might include noisy and redundant information and increase computational times of automated EEG decoding algorithms. To reduce the signal-to-noise ratio, improve accuracy and reduce computational time, one may combine channel selection with feature extraction and dimensionality reduction. However, as EEG signals present high inter-subject variability, we introduce a novel algorithm for subject-independent channel selection through representation learning of EEG recordings. The algorithm exploits channel-specific 1D-CNNs as supervised feature extractors to maximize class separability and reduces a high dimensional multi-channel signal into a unique 1-Dimensional representation from which it selects the most relevant channels for classification. The algorithm can be transferred to new signals from new subjects and obtain novel highly informative trial vectors of controlled dimensionality to be fed to any kind of classifier.

\end{abstract}
\section{Introduction}
\label{sec:1}
ElectroEncephaloGraphy (EEG) is a non-invasive system that places external electrodes along the scalp, and measures voltage fluctuations resulting from ionic current within the neurons of the brain. This powerful technology finds applications in several fields \cite{lehmann2007application, cai2016pervasive, albert2016automatic, fathima2019enhanced}, all relying on EEG technology because of its high portability, relative low cost, high temporal resolution and few risk to users. All these aspects could drive the development of portable devices to assist impaired patients, medical systems for early detection and personalized treatment of neurological diseases \cite{faul2010dynamic, moctezuma2020eeg} and useful tools for clinical decision making.
However, irrespectively of the specific application, to achieve good performance in EEG decoding, patients are required to wear several electrodes (\textit{channels}). Applying a large number of channels presents several drawbacks: it could (i) include noisy and redundant signals affecting analysis; (ii) induce longer preparation times in clinical studies, that become longer and more expensive; (iii) reduce the portability and convenience of EEG-based assistive technologies such as BCI devices; (iv) lead to higher computational time in the automated processing of signal data for classification purposes and (v) strongly affect the performance of any Statistical or Machine Learning approach.
The development of effective channel selection algorithms is one of the most relevant strategies to overcome all the aforementioned issues at once \cite{lan2006salient}. However, EEG data is known to be highly non-stationary and subject variant, i.e. there exists inter-subject variability wherein the best-performing channels for one user may not be the same for another user.
Unsurprisingly, most existing literature focuses on \textit{subject-dependent} channel selection strategies, where the specific subset of channels or the ranking of channels' importance is performed for each subject independently \cite{lan2006salient, arvaneh2011optimizing, handiru2016optimized, fathima2019enhanced, moctezuma2020eeg, he2013channel, wei2011binary} or at a group level, where the same set of channels is selected across an entire group of subjects, but the selection remains valid  - and is evaluated - for new signals from the same group of subjects \cite{yang2012channel, arvaneh2011optimizing, ong2006selection}. Some attempts of cross-subject selection exist  \cite{schroder2005robust, handiru2016optimized, atum2019comparison}, but either performance is quite low or the number of selected channels high, or the methods tailored to specific EEG study paradigms.
Despite the complexity of the task, it has become apparent that crucial medical applications call for automated EEG decoding via different Statistical and Machine Learning approaches \cite{gemein2020machine}, that in turn need strategies to reduce signal-to-noise ratio and improve classification accuracy of EEG recordings.
To this end, besides - and combined to - effective channel selection, feature extraction from the selected raw signals is another indispensable step of EEG decoding tasks, to reduce dimensionality and foster classification performance \cite{lan2006salient}. Feature-based methods, where typically handcrafted and \textit{a priori} selected features represent the data, are the most widely adopted, but rely on domain expertise of the researcher \cite{gemein2020machine}. Conversely, \textit{End-to-End} decoding methods accept raw or minimally preprocessed data as input and automatically extract features that are relevant for the specific classification task. However, they are oftentimes less interpretable and very expensive in computation and memory requirement terms, due to the complexity of the underlying models \cite{gemein2020machine}.\\
Finding the most effective and generalizable combination of features extraction and channel selection can balance both needs for performance and convenience of potential real-life applications of EEG-based systems and overcome all the aforementioned complexities in automated EEG analysis implementations.
For this reason, in this study we focus on the development of an algorithm that combines a Representation Learning (RL) method for signals' feature extraction (namely 1D Convolutional Neural Network - 1DCNN), with a novel channel selection method.  
Indeed, we present here \texttt{ERNEST}, EEG \texttt{E}mbedde\texttt{R} a\texttt{N}d chann\texttt{E}l \texttt{S}elec\texttt{T}or: a robust, generalizable, simple and lightweight End-to-End algorithm for EEG channel selection and signal decoding.\\
As a first proof of concept of our algorithm we decided to focus on a specific EEG classification setting, namely alcoholism detection. This case study may serve as an interesting experiment due to the relevance of this problem in the medical domain, as testified by the several previous studies that tried to tackle this classification task by means of different channel selection approaches. In \cite{bavkar2021optimal} a thorough review of previous attempts is presented. However, these previous examples, most of which with very high accuracy performances, mainly focus on selecting channels for classification in a group-dependent fashion. A subject-agnostic channel selection in this context could instead be a valuable component of a clinical decision support system for the early detection of predisposition to alcoholism on new patients. 

\vspace{-10pt}
\section{Methods}

\begin{figure*}[!t]
    \centering
    \includegraphics[width=1.6\columnwidth]{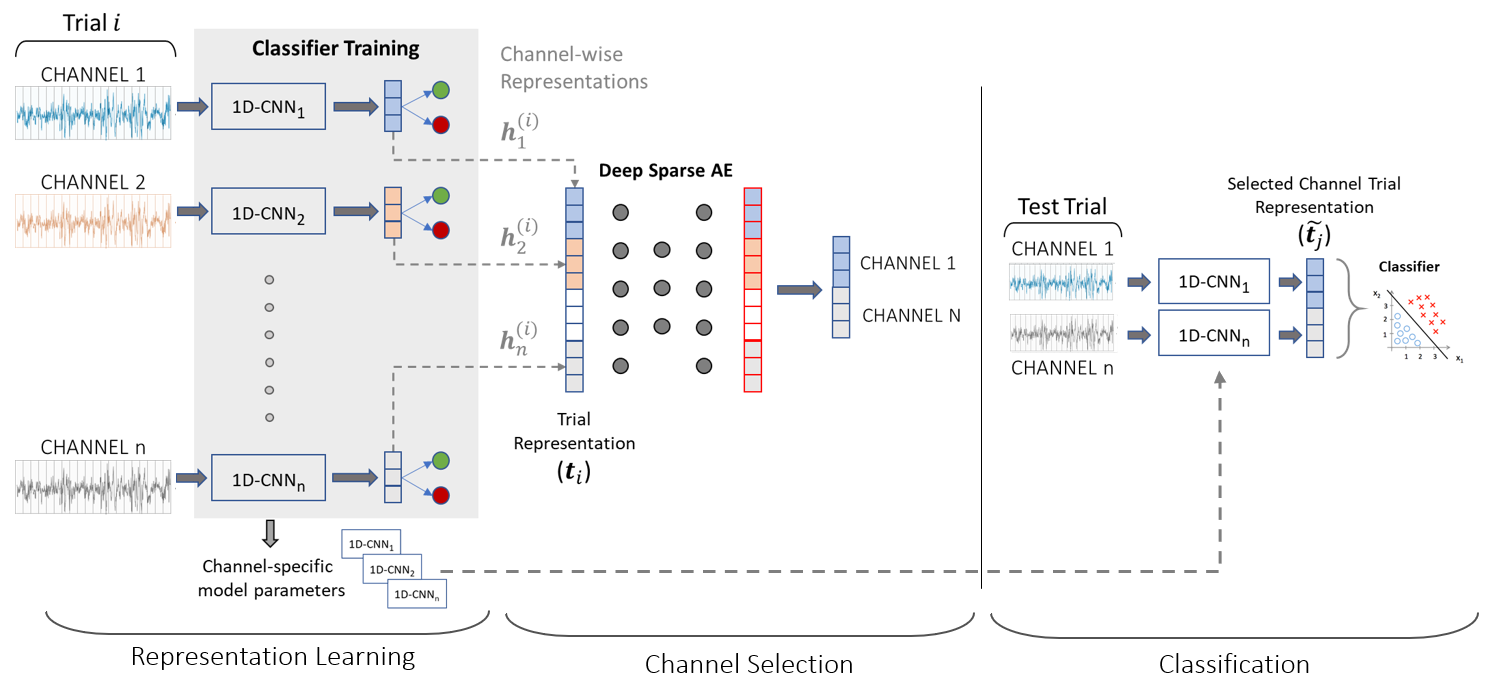}
    \caption{\texttt{ERNEST} process flow}
    \label{fig:process}
\end{figure*}

The aim of this paper is to present \texttt{ERNEST}, an End-to-End RL-based algorithm to reduce signal dimensionality in a channel-wise fashion and select the most relevant electrodes to perform classification across subjects. In other words, regardless of the subjects employed during the training phase, the algorithm should learn an effective mapping from the multi-channel EEG signals to a reduced 1-dimensional representation (retaining information from the most relevant channels only) that provide a sufficient accuracy on new signals from any patient. 
\texttt{ERNEST} is composed of modules tailored to address different parts of the process. Indeed, the algorithm follows a multi-step process as depicted in Figure \ref{fig:process}. Note that the algorithm works on set of signal recordings that we will here define generally \textit{trials}, i.e. fixed length multi-dimensional signals recorded simultaneously from several electrodes: one trial might be one stimulus of a visual or auditory Event-Related-Potential (ERP), or an epoch within a Motor Imagery (MI) experiment, or a chunk of recordings for patients' classification.
\vspace{-5pt}

\subsection{Channel-wise feature extraction} \label{feature_extr} First of all, \texttt{ERNEST} exploits a 1D-CNN to automatically extract channel-specific features from signals. To do that, it parametrizes a set of channel-specific embedding models optimized to separate between different classes of trials. The aforementioned classes might represent the health status of a patient, the onset of an epileptic seizure, the different responses to stimuli in an ERP trial or the imaginated motor task in a MI experiment. As mentioned, a notable aspect of the algorithm is the lack of signal preprocessing because of the end-to-end nature of 1DCNNs.\\
In particular, we can consider each 1DCNN as composed of an \textit{encoder} and a subsequent \textit{classifier}. The encoder maps the signals from the $J$-dimensional input space, into an $M$-dimensional embedding space, where $M<J$. The whole model is parametrized with supervised training to classify the signals in the classes of interest. After training, the embedded $M$-dimensional vectors from each of the $C$ channels are extracted from the \textit{encoder}, and the algorithm builds a unique 1D representation of each trial by concatenating the $C$ embeddings into a trial vector $\textbf{t} \in \rm I\!R^{1\times(M\times C)}$.\\
The choice of training several parallel models was led by the idea that parametrizing in a supervised channel-wise fashion would generate rich embeddings for all channels and better capture the independent relationship each of those electrode locations has with the target. However, the information lying in the combination of several recording sources and their inter-relationships is not lost, as it is considered in the subsequent Autoencoder ensemble-based Channel Selection (CS) module applied to trial vectors, as described in the following.

\vspace{-5pt}
\subsection{Channel selection} The CS module relies on a supervised  filtering (i.e. classifier-agnostic) Feature Selection method developed in \cite{massi2019minority}. This method exploits an Ensemble of Deep Sparse AutoEncoders (DSAEE) to select the most relevant features to discriminate between classes in a binary setting. 
The algorithm is adapted to select channels instead of single features and its AE-based nature is suited to consider complex non-linear inter-channel relationships while performing selection. In particular, after training of the Feature Extraction module, all training trial vectors $\textbf{t}$ together form the matrix $\textbf{T} \in \rm I\!R^{N \times (CM)}$, where $CM$ is the product between the number of channels and the embedding dimension, that will be used to select the most relevant channels via DSAEE. We supply $\textbf{T}$ to the algorithm that (repeating for each of the $B$ ensemble components) (i) builds the test set ($\textbf{T}_{test}$) by sampling a subset with the same number $L$ of observations from each class, (ii) builds the training set $\textbf{T}_{train} = \{ \textbf{t}_{i}|y_{i}=0, \forall i\notin \textbf{T}_{test} \}$ with all observations from one of the classes ($y_{i}=0$) not included in the test set. Then (iii) it trains the DSAE to reconstruct $\textbf{T}_{train}$, and collects the RE on $\textbf{T}_{test}$ as the Squared Error between the input and its reconstruction. The final RE matrix $\textbf{R} \in \rm I\!R^{2BL \times CM}$ is the final output, $2BL$ being the total number of tested observations by the ensemble of models, and the features represent the testing RE committed by the DSAE on the $C$ concatenated $M$-dimensional channel-specific vectors. The RE \textit{by class} for channel $c$ is then computed summing over the $M$-dimensional chunk in \textbf{R} associated with $c$ and taking the mean of the $BL$ trials of the class:
$$RE^{(c)}_{class}=\frac{1}{L}\sum_{\{i|y_{i}=class\}}\sum_{m=1}^{M}r^{(c)}_{i,m}$$
where the $class$ subscript represents the levels of $\textbf{y} \in \{0,1\}$, and each vector $\textbf{r}^{(c)}_{i} \in \rm I\!R^{M}$ is the subset vector of RE associated to channel $c$ for each trial $i$. The final metric to evaluate channels' relevance is the difference in RE between classes, i.e. $\Delta RE^{(c)} = RE^{(c)}_{1} - RE^{(c)}_{0}$. We rank order channels on such metric, and we select the top $k$ as the most salient to discriminate between the two classes of trials.

\vspace{-5pt}
\subsection{New trials representation through selected-channel dimensionality reduction}
Once the top $K$ most relevant channels have been selected, \texttt{ERNEST} tackles the classification of new trials by exploiting the pretrained models associated to each selected channel. Indeed, at the end of the training phase, each channel is associated with an embedding function $\phi_{c}$ parametrized with supervision to maximize class separability.
Given an unseen test set of (trial, target) pairs, for each trial the algorithm filters on the selected $K$ channels, and concatenates the representations obtained applying $\phi_{k}$ to the corresponding signal, with $k=(1,...,K)$. By doing that, we strongly reduce the dimensionality of the problem by transforming a high-dimensional multi-channel representation of each trial into a single vector of controlled size that is highly informative w.r.t. the task at hand. Indeed, we end up with the matrix $\textbf{A} \in \rm I\!R^{P \times (KM)}$, where $P$ is the number of new trials, $K$ is the number of selected relevant channels and $M$ is the representation dimension. As mentioned, each trial is associated to a target, that constitutes the vector $\textbf{y}_{new} \in \rm I\!R^{P}$ associated to $\textbf{A}$. 
The matrix $\textbf{A}$ and the associated target vector can then be supplied to any classifier.

\section{Case study application: Early detection of alcoholism predisposition}
Besides the general description of the algorithm, in this work we aim at testing its potential and proving some details of its conceptualization by running an experiment on real data.\\
The dataset we use is a large public EEG database, available through UCI Machine Learning Repository\footnote{https://archive.ics.uci.edu/ml/datasets/EEG+Database}. This dataset was developed to examine genetic predisposition, through EEG signals, to alcoholism. To elicit the Event-Related Potential (ERP), a modified delayed Visually Evoked Potential (VEP) matching-to-sample task was used, in which two picture stimuli appeared in succession: a first picture stimulus (S1) was followed by a second stimulus (S2) either matching or non-matching the first picture.
The database includes 122 subjects, equally splitted between alcoholic and controls. Each subject completed 120 trials. The signal acquisition is performed according to the 10–20 International System with 64 electrodes placed on the scalps of the subjects and recordings were sampled at 256 Hz (3.9-msec epoch) for 1 second.

\vspace{-10pt}
\subsection{Experimental and Implementation Details}
\begin{figure}[!t]
\centering
    \includegraphics[width=\columnwidth]{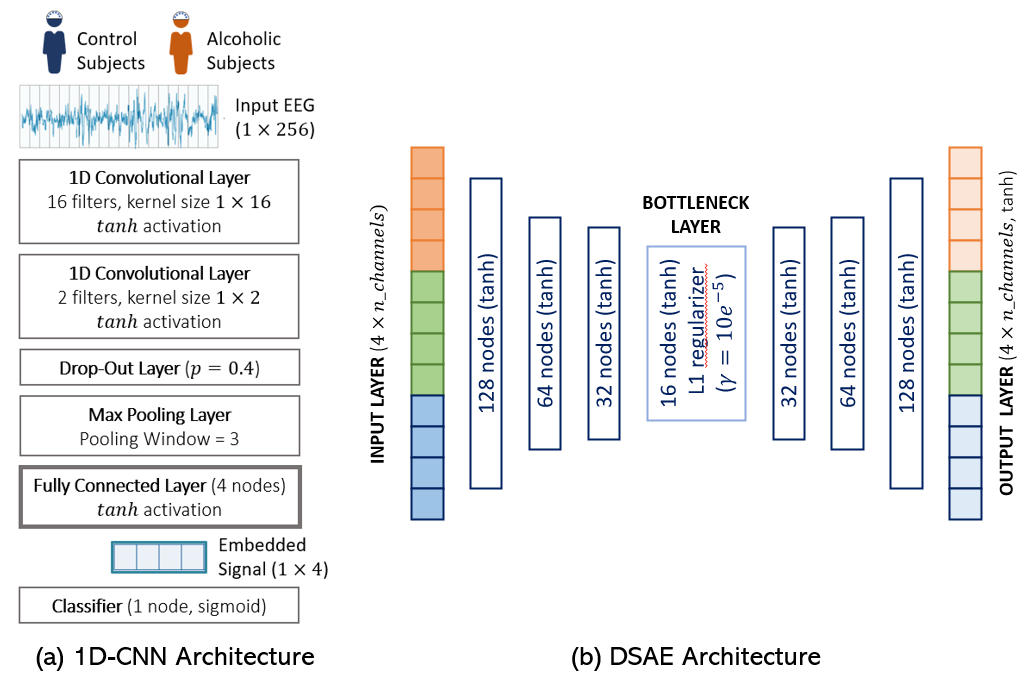}
    \caption{(a) Architectural details of the 1D-CNNs employed for the described experiment. (b) DSAE components' architectural details. DSAEs are exploited for channel selection from embedded trial vectors.}
    \label{fig:architect}
\vspace{-10pt}
\end{figure}

Considering that VEP are generally known to grant higher accuracies compared to other paradigms, for this proof of concept experiment we focused on a more complex classification objective by trying to classify trials (the class is determined by the subject associated to each trial) on the basis of the brain signals produced as response to the first visual stimulus (S1) only.
We splitted the 122 subjects in training (102 subjects, equally divided between alcoholics and controls) and test group (20 new subjects, 50\% alcoholics and 50\% controls). The former group with all its trials was exploited for channel-wise 1DCNNs training and channel selection, while the latter was supplied to the algorithm to test the cross-subject classification performance.
To perform the channel-wise embedding of the EEG recordings we had to train several channel-specific models. We opted for a shallow 1D-CNN and the specific architectural details are reported in Figure \ref{fig:architect}.(a). Notably, signals from each channel were reduced to the very low dimension of $M=4$. Hyperparameters were chosen by randomly sampling 10,000 training signals irrespectively of the channel - to make the tuning generalizable across electrodes - equally splitted between classes, and performing random search of the best combination. After setting hyperparameters, each channel-specific 1D-CNN was trained for 200 epochs with a batch size of 1,000 signals.
For what concerns the channel selection module, details are reported in Figure \ref{fig:architect}.(b). Each DSAE model in the Ensemble (30 components in total) was trained for 300 epochs with a batch size of 500 training trials.
The whole algorithm was implemented in Python 3.7, exploiting Keras framework with Tensorflow backend and scikit-learn. The code is available upon request to allow for reproducibility of results.
To evaluate the performance on test set we adopted several classifiers, namely Support Vector Machines (SVM), Random Forests (RF) and Logistic Regression (LR). We evaluated whether the channel reduction would impact the performance of the classifiers by first trying to classify trials using all 61 channels (after their embedding via 1DCNNs and transformation into trial vectors) and then with smaller subsets of $K = \{30, 20, 15, 10, 5\}$ most relevant channels. The resulting performance was measured using the Area Under the ROC curve (AUROC) and Accuracy metrics by cross-validating 10 times. Whenever possible, we will include in the following results the empirical comparison to the limited literature pursuing cross-subject channel selection on this data. However, note that those results are based on the simpler classification on S2 stimuli as well. Nonetheless, with this first case study application of \texttt{ERNEST} we are more interested in proving the underlying hypotheses of the algorithm, while attaining a satisfactory classification performance. Indeed, we include in our experiment an ablation study that aims at verifying the value added by the DSAEE-based CS Module. To do that, we rank order channels on the basis of the accuracy of channel-specific 1D-CNNs in classifying alcoholics and controls during Feature Extraction Module training. Then, we build the trial vectors by concatenating the top K (in terms of accuracy) channel-specific embeddings, and we feed them to the group of classifiers. 
\begin{table}[]
\centering
\label{tab:exp_b}
\resizebox{0.65\columnwidth}{!}{%
\begin{tabular}{lllllll}
\hline
\multicolumn{7}{c}{\textbf{Detection Performance}} \\ \hline
 & \multicolumn{2}{c}{SVM} & \multicolumn{2}{c}{RF} & \multicolumn{2}{c}{LR} \\ \cline{2-7} 
 & \multicolumn{6}{c}{AUROC} \\ \cline{2-7} 
K  & Mean & Std & Mean & Std & Mean & Std \\
61 & \textbf{0.905} & \textbf{0.013} & 0.826 & 0.018 & 0.890 & 0.014 \\
30 & \textbf{0.895} & \textbf{0.018} & 0.834 & 0.018 & 0.841 & 0.019 \\
20 & \textbf{0.879} & \textbf{0.020} & 0.835 & 0.018 & 0.822 & 0.015 \\
15 & \textbf{0.862} & \textbf{0.016} & 0.842 & 0.022 & 0.799 & 0.013 \\
10 & \textbf{0.872} & \textbf{0.021} & 0.846 & 0.028 & 0.800 & 0.018 \\
5 & \textbf{0.858} & \textbf{0.018} & 0.808 & 0.032 & 0.759 & 0.021 \\ \cline{2-7}
 & \multicolumn{6}{c}{ACCURACY} \\ \cline{2-7} 
K & Mean & Std & Mean & Std & Mean & Std \\
61 & \textbf{0.807} & \textbf{0.015} & 0.736 & 0.018 & \textbf{0.808} & \textbf{0.014} \\
30 & \textbf{0.816} & \textbf{0.017} & 0.766 & 0.020 & 0.787 & 0.016 \\
20 & \textbf{0.798} & \textbf{0.023} & 0.766 & 0.011 & 0.765 & 0.019 \\
15 & \textbf{0.793} & \textbf{0.024} & 0.763 & 0.017 & 0.740 & 0.021 \\
10 & \textbf{0.786} & \textbf{0.025} & 0.766 & 0.019 & 0.726 & 0.032 \\
5 & \textbf{0.762} & \textbf{0.015} & 0.733 & 0.020 & 0.716 & 0.017 \\ \cline{2-7}
\end{tabular}%
}
\caption{Trial classification results with SVM, RF and LR classifiers in terms of AUROC and Accuracy}
\end{table}
\vspace{-10pt}

\begin{table}[]
\centering
\vspace{-5pt}
\label{tab:exp_b_ablation}
\resizebox{0.65\columnwidth}{!}{%
\begin{tabular}{lrrrrrr}
\hline
\multicolumn{7}{c}{Detection Performance w/o DSAEE CS Module} \\ \hline
 & \multicolumn{2}{c}{SVM} & \multicolumn{2}{c}{RF} & \multicolumn{2}{c}{LR} \\ \cline{2-7} 
 & \multicolumn{6}{c}{AUROC} \\ \cline{2-7} 
K & \multicolumn{1}{l}{Mean} & \multicolumn{1}{l}{Std} & \multicolumn{1}{l}{Mean} & \multicolumn{1}{l}{Std} & \multicolumn{1}{l}{Mean} & \multicolumn{1}{l}{Std} \\
30 & \textbf{0.877} & \textbf{0.017} & 0.792 & 0.022 & 0.833 & 0.019 \\
20 & \textbf{0.859} & \textbf{0.020} & 0.783 & 0.023 & 0.826 & 0.016 \\
15 & \textbf{0.815} & \textbf{0.020} & 0.762 & 0.021 & 0.806 & 0.016 \\
10 & \textbf{0.801} & \textbf{0.022} & 0.758 & 0.024 & 0.793 & 0.022 \\
5 & \textbf{0.778} & \textbf{0.020} & 0.754 & 0.026 & 0.773 & 0.021 \\ \cline{2-7} 
 & \multicolumn{6}{c}{ACCURACY} \\ \cline{2-7} 
K & \multicolumn{1}{l}{Mean} & \multicolumn{1}{l}{Std} & \multicolumn{1}{l}{Mean} & \multicolumn{1}{l}{Std} & \multicolumn{1}{l}{Mean} & \multicolumn{1}{l}{Std} \\
30 & \textbf{0.786} & \textbf{0.020} & 0.709 & 0.019 & 0.756 & 0.023 \\
20 & \textbf{0.768} & \textbf{0.025} & 0.709 & 0.024 & 0.748 & 0.024 \\
15 & \textbf{0.744} & \textbf{0.025} & 0.687 & 0.019 & 0.724 & 0.020 \\
10 & \textbf{0.730} & \textbf{0.026} & 0.695 & 0.024 & 0.714 & 0.022 \\
5 & \textbf{0.713} & \textbf{0.021} & 0.689 & 0.024 & 0.704 & 0.021 \\ \cline{2-7} 
\end{tabular}%
}
\caption{Trial classification results with SVM, RF and LR classifiers (AUROC and Accuracy) without ERNEST Channel Selection module. Performance for K=61 is the one reported in Table 1.}
\end{table}

\subsection{Results}
\label{res}
Results for this experiment are reported in Table 1. This experimental setting is lightly comparable to \cite{ong2006selection, silva2020classification} and benchmark algorithms therein, even though performance measurements and data splitting criteria are not always clear from the original papers. Our algorithm obtains a satisfactory accuracy, and a very high AUROC performance, indicating a great precision in identifying the positive class (i.e. \textit{alcoholics}). Our best classifier (SVM) with only 5 electrodes surpasses the performance in \cite{ong2006selection} with 4 channels ($75.13\%$ average accuracy). However, in this work the authors exploit PCA for channel selection applied to the whole dataset, and the lack of information on splitting criteria or performance standard deviations suggest that they are reporting training accuracy measures, which are overestimated compared to our test values. The average accuracy reported more recently in \cite{silva2020classification} with 11 channels ($\sim 93\% \pm 3.3$ with the best proposed approach and SVM classifier) is therein defined as the state-of-the-art on this data. Their performance is higher compared to ours with a similar number of channels. However, note that their performance reflects the easier task including S2 stimuli, and they perform channel selection evaluating the mean-variance of each channel for all subjects in the dataset before proceeding with feature extraction and classification, therefore their selection is not comparable to our subject-agnostic approach. 
\begin{figure}[t]
\centering
    \includegraphics[width=\columnwidth]{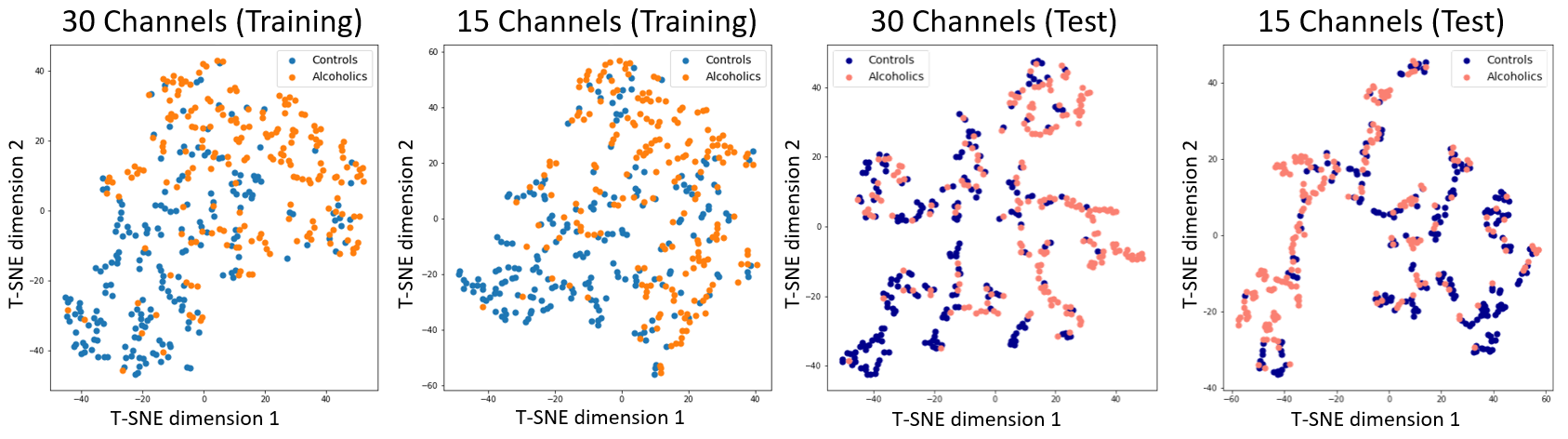}
    \caption{t-SNE plots of trial vectors with $K=\{30,15\}$ of a sample of training and test trials.}
    \label{fig:tsne}
\end{figure}
In Figure \ref{fig:tsne} we plot the 2D t-SNE representations of 200 sampled trial vectors, that seem consistently separable in both training and test trials from new subjects.
Finally, in Table 2 we report the results of the ablation study without the use of the AE-based CS Module. SVM remains the best classifier for this task, but the performance drop can be recognized consistently across all classifiers. Moreover, note that while in both cases the reduction in number of channels leads to a decrease in performance, when including the CS module the drop is significantly smaller: from $0.905$ mean AUROC with $K=61$ to $0.858$ mean AUROC for \texttt{ERNEST} with CS module and $0.778$ mean AUROC for \texttt{ERNEST} w/o CS module for $K=5$. This testifies for this module's capability to effectively consider inter-channel relationships when selecting the relevant channels for classification, identifying the smallest most informative subset.

\vspace{-5pt}
\section{Discussion and Conclusions}

In this work we proposed \texttt{ERNEST}, an end-to-end RL-based algorithm for EEG feature extraction and channel selection. We empirically demonstrated its potential by running a case study experiment on real data in which \texttt{ERNEST} yielded good performance results, supporting the underlying hypothesis that grounded its design. Besides classification performance - which is hard to benchmark with other existing approaches because of the multi-step and multi-purpose nature of its building blocks - the algorithm presents several additional advantages one should consider. First of all, the supervised channel-specific 1D-CNNs of its Feature Extraction module seem to capture the needed information for class separation into very small signal embeddings (4 dimensions per channel only): this allows users to avoid cumbersome and knowledge-intensive data preprocessing, while keeping a manageable dimensionality of the resulting concatenated trial vectors, that can be easily exploited for any following ML or statistical task. Larger embedding dimensions $M$ might have granted even more competitive accuracy results, but would have reduced this added value. Of course, a limit of the algorithm resides in the need for larger amounts of data to improve in generalizability and obtain effective representations. We tried to limit the impact of this aspect by keeping the models in \texttt{ERNEST} RL module shallow and opting for CNNs, that by weight-sharing reduce the number of parameters to learn. Indeed, in the described case study each CNN have only $\sim 1000$ parameters. Moreover, the novel approach of parametrizing channel-specific models increases the distance between each channel's embedding and enriches their representations, forcing each 1D-CNN to learn precise representations of each electrode and their role in determining the target, reducing the risk of losing precious information in the noise of recording simultaneous signals. \texttt{ERNEST} combines these independently learnt representation with a CS module (i.e. the DSAEE), that is meant to re-capture the complex relationships between channels that might per se contain useful information to separate classes. Of course this approach negatively impacts computational time, but a further advantage of parametrizing separate models is the possibility - after channel selection - to store and apply only a smaller subset of them for the embedding of new trials, reducing each high-dimensional multi-channel trial observation into unique vector of controlled size that is highly informative w.r.t. the task at hand. This aspect qualifies \texttt{ERNEST} as a smart supervised dimensionality reduction technique, counterbalancing the time complexity of the training phase with an extremely fast transformation of new trials: this is indeed the step where its proven cross-subject generalizability provides value to real-life applications on new subjects.\\ 
We applied \texttt{ERNEST} to the very specific clinical task of early diagnosis of predisposition to alcoholism, but several other medical fields share the need for a powerful cross-subject channel selection method. Moreover, a limit of this preliminary proof of concept study on the algorithm is the application on a unique EEG recording paradigm (i.e. VEP), while other paradigms may present even stronger inter-subject variability or the need for inclusion of prior domain knowledge in channel selection, which would require adjustments to the present algorithm. Nonetheless, the promising performance yielded in this first application makes it an interesting starting point for future research to extend its framework to different EEG applications and paradigms.

\bibliographystyle{IEEEbib}
\bibliography{refs}

\begin{thebibliography}{10}

\bibitem{lehmann2007application}
C.~Lehmann, T.~Koenig, V.~Jelic, L.~Prichep, R.~E. John, L.-O. Wahlund,
  Y.~Dodge, and T.~Dierks,
\newblock ``Application and comparison of classification algorithms for
  recognition of alzheimer's disease in electrical brain activity (eeg),''
\newblock {\em Journal of neuroscience methods}, vol. 161, no. 2, pp. 342--350,
  2007.

\bibitem{cai2016pervasive}
H.~Cai, X.~Sha, X.~Han, S.~Wei, and B.~Hu,
\newblock ``Pervasive eeg diagnosis of depression using deep belief network
  with three-electrodes eeg collector,''
\newblock in {\em 2016 IEEE International Conference on Bioinformatics and
  Biomedicine (BIBM)}. IEEE, 2016, pp. 1239--1246.

\bibitem{albert2016automatic}
B.~Albert, J.~Zhang, A.~Noyvirt, R.~Setchi, H.~Sjaaheim, S.~Velikova, and
  F.~Strisland,
\newblock ``Automatic eeg processing for the early diagnosis of traumatic brain
  injury,''
\newblock in {\em 2016 World Automation Congress (WAC)}. IEEE, 2016, pp. 1--6.

\bibitem{fathima2019enhanced}
S.~Fathima and S.~K. Kore,
\newblock ``Enhanced differential evolution-based eeg channel selection,''
\newblock in {\em Symposium on Machine Learning and Metaheuristics Algorithms,
  and Applications}. Springer, 2019, pp. 162--174.

\bibitem{faul2010dynamic}
S.~D. Faul,
\newblock ``Dynamic channel selection to reduce computational burden in seizure
  detection,''
\newblock in {\em 2010 Annual International Conference of the IEEE Engineering
  in Medicine and Biology}. IEEE, 2010, pp. 6365--6368.

\bibitem{moctezuma2020eeg}
L.~A. Moctezuma and M.~Molinas,
\newblock ``Eeg channel-selection method for epileptic-seizure classification
  based on multi-objective optimization,''
\newblock {\em Frontiers in Neuroscience}, vol. 14, pp. 593, 2020.

\bibitem{lan2006salient}
T.~Lan, D.~Erdogmus, A.~Adami, M.~Pavel, and S.~Mathan,
\newblock ``Salient eeg channel selection in brain computer interfaces by
  mutual information maximization,''
\newblock in {\em 2005 IEEE Engineering in Medicine and Biology 27th Annual
  Conference}. IEEE, 2006, pp. 7064--7067.

\bibitem{arvaneh2011optimizing}
M.~Arvaneh, C.~Guan, K.~K. Ang, and C.~Quek,
\newblock ``Optimizing the channel selection and classification accuracy in
  eeg-based bci,''
\newblock {\em IEEE Transactions on Biomedical Engineering}, vol. 58, no. 6,
  pp. 1865--1873, 2011.

\bibitem{handiru2016optimized}
V.~S. Handiru and V.~A. Prasad,
\newblock ``Optimized bi-objective eeg channel selection and cross-subject
  generalization with brain--computer interfaces,''
\newblock {\em IEEE Transactions on Human-Machine Systems}, vol. 46, no. 6, pp.
  777--786, 2016.

\bibitem{he2013channel}
L.~He, Y.~Hu, Y.~Li, and D.~Li,
\newblock ``Channel selection by rayleigh coefficient maximization based
  genetic algorithm for classifying single-trial motor imagery eeg,''
\newblock {\em Neurocomputing}, vol. 121, pp. 423--433, 2013.

\bibitem{wei2011binary}
Q.~Wei and Y.~Wang,
\newblock ``Binary multi-objective particle swarm optimization for channel
  selection in motor imagery based brain-computer interfaces,''
\newblock in {\em 2011 4th International conference on biomedical engineering
  and informatics (BMEI)}. IEEE, 2011, vol.~2, pp. 667--670.

\bibitem{yang2012channel}
J.~Yang, H.~Singh, E.~L. Hines, F.~Schlaghecken, D.~D. Iliescu, M.~S. Leeson,
  and N.~G. Stocks,
\newblock ``Channel selection and classification of electroencephalogram
  signals: an artificial neural network and genetic algorithm-based approach,''
\newblock {\em Artificial intelligence in medicine}, vol. 55, no. 2, pp.
  117--126, 2012.

\bibitem{ong2006selection}
K.-M. Ong, K.-H. Thung, C.-Y. Wee, and R.~Paramesran,
\newblock ``Selection of a subset of eeg channels using pca to classify
  alcoholics and non-alcoholics,''
\newblock in {\em 2005 IEEE Engineering in Medicine and Biology 27th Annual
  Conference}. IEEE, 2006, pp. 4195--4198.

\bibitem{schroder2005robust}
M.~Schr{\"o}der, T.~N. Lal, T.~Hinterberger, M.~Bogdan, N.~J. Hill,
  N.~Birbaumer, W.~Rosenstiel, and B.~Sch{\"o}lkopf,
\newblock ``Robust eeg channel selection across subjects for brain-computer
  interfaces,''
\newblock {\em EURASIP Journal on Advances in Signal Processing}, vol. 2005,
  no. 19, pp. 174746, 2005.

\bibitem{atum2019comparison}
Y.~Atum, M.~Pacheco, R.~Acevedo, C.~Tabernig, and J.~B. Manresa,
\newblock ``A comparison of subject-dependent and subject-independent channel
  selection strategies for single-trial p300 brain computer interfaces,''
\newblock {\em Medical \& biological engineering \& computing}, vol. 57, no.
  12, pp. 2705--2715, 2019.

\bibitem{gemein2020machine}
L.~A. Gemein, R.~T. Schirrmeister, P.~Chrabaszcz, D.~Wilson, J.~Boedecker,
  A.~Schulze-Bonhage, F.~Hutter, and T.~Ball,
\newblock ``Machine-learning-based diagnostics of eeg pathology,''
\newblock {\em NeuroImage}, p. 117021, 2020.

\bibitem{bavkar2021optimal}
S.~Bavkar, B.~Iyer, and S.~Deosarkar,
\newblock ``Optimal eeg channels selection for alcoholism screening using emd
  domain statistical features and harmony search algorithm,''
\newblock {\em Biocybernetics and Biomedical Engineering}, vol. 41, no. 1, pp.
  83--96, 2021.

\bibitem{massi2019minority}
M.~C. Massi, F.~Ieva, F.~Gasperoni, and A.~M. Paganoni,
\newblock ``Feature selection for imbalanced data with deep sparse autoencoders
  ensemble,''
\newblock {\em arXiv preprint arXiv:2103.11678}, 2021.

\bibitem{silva2020classification}
F.~H. Silva, A.~G. Medeiros, E.~F. Ohata, and P.~P. Reboucas~Filho,
\newblock ``Classification of electroencephalogram signals for detecting
  predisposition to alcoholism using computer vision and transfer learning,''
\newblock in {\em 2020 IEEE 33rd International Symposium on Computer-Based
  Medical Systems (CBMS)}. IEEE, 2020, pp. 126--131.

\end{thebibliography}

\end{document}